\documentstyle[twoside,fleqn,espcrc2,epsf]{article}

\newcommand{\AmS}{{\protect\the\textfont2
  A\kern-.1667em\lower.5ex\hbox{M}\kern-.125emS}}
                               
          \def\s{\sigma}                                
           \def\l{\lambda}                    
\def\D{\Delta}           \def\W{\Omega}            \def\G{\Gamma}       
\def\k{{\bf k}}          \def\r{{\bf r}}                 
                           
\def\S{{\cal S}}                   \def\Hso{{\cal H}_{\rm so}}  
                  
\def\RT{R_{\rm T}}        \def\GT{G_{\rm T}}   
\def\lso{\l_{\rm so}}    \def\lS{\l_{\cal S}}        
       
\def\Vb{V_{\rm }}        \def\IA{I_{\rm AP}}       \def\IF{I_{\rm P}}   
\def\/{\over}            \def\us{\uparrow}         \def\ds{\downarrow}  
\def\dmu{\delta\mu}              \def\vF{v_{\rm F}}
\def\[{\left[}           \def\]{\right]}           
\def\({\left(}           \def\){\right)}           
\def\<{\langle}          \def\>{\rangle}           \def\Ekp{E_{\bf k'}} 
             \def\Ek{E_{\bf k}}   
               
\def\xk{\xi_{\k}}
                   
\def\tS{\tau_{s}}        \def\tsf{\tau_{\rm sf}}
\def\GS{\G_{s}}          \def\GSN{\G_s^{\rm N}}    \def\nimp{n_{\rm i}}
\def\timp{\tau_{\rm imp}}\def\Vimp{V_{\rm imp}}    \def\tsf{\tau_{\rm sf}}

\def\Xy{\chi(T)}   
             
\def\dt{\partial t}             
               
\def\d{\partial}         \def\h{\hbar}

\title{Spin-relaxation and magnetoresistance in FM/SC/FM tunnel junctions}

\author{S. Takahashi\address[MCSD]{Institute for Materials Research,
       Tohoku University, Sendai 980-8577, Japan}
T. Yamashita\addressmark[MCSD],
H. Imamura\address{Graduate School of Information Sciences,
       Tohoku University, Sendai 980-8579, Japan}, and
S. Maekawa\addressmark[MCSD] }

\begin{document}

\begin{abstract}
The effect of spin relaxation on tunnel magnetoresistance (TMR) in
a ferromagnet/superconductor/ferromagnet (FM/SC/FM) double tunnel
junction is theoretically studied.  The spin accumulation in SC is determined
by balancing of the spin-injection rate and the spin-relaxation rate.
In the superconducting state, the spin-relaxation time $\tS$ becomes longer
with decreasing temperature, resulting in a rapid increase of TMR.
The TMR of FM/SC/FM junctions provides a useful
probe to extract information about spin-relaxation in superconductors.
\end{abstract}

\maketitle


Spin-polarized tunneling plays an important role in the spin-dependent
transport of magnetic nanostructures \cite{meservey}.  
The spin-polarized electrons injected from ferromagnets (FM)
into nonmagnetic metals (NM) such as a normal metal, semiconductor,
and superconductor creates a nonequilibrium spin polarization in NM
 \cite{johnson,varet,vanwees,johnsonS,vasko,dong,daibo}.
The efficient spin injection and weak spin-relaxation during
transport are required for practical applications.  A number of
experiments for observing the spin relaxation time $\tS$ in SCs has
been reported by using a spin-injection device \cite{johnsonS}
and by the conduction electron spin resonance \cite{vier,nemes}.

A double tunnel junction FM/SC/FM containing superconductor (SC)
sandwiched between two FMs is a unique system to investigate
nonequilibrium phenomena caused by spin injection, especially the
magnetoresistive effects by competition between superconductivity
and spin accumulation \cite{takahashiPRL,takahashiJAP,takahashiPC}.
The pronounced magnetoresistance effects is brought about by a long
spin relaxation time $\tS$ in SC, which corresponds to a long
spin-diffusion length.
In this article, we take into account the coherence effect of
superconductivity on the spin-relaxation due to spin-orbit scattering
by impurities \cite{yafet}, and demonstrate that the tunnel
magnetoresistance (TMR) of the FM/SC/FM junction exhibits a large
enhancement due to the increase of $\tS$ in the superconducting state.


We consider a FM/SC/FM double tunnel junction.
The left and right electrodes are made of a ferromagnet, and the
central one is a superconductor with thickness $d$ smaller than the
spin-diffusion length $\lS$.  The magnetizations of FMs are aligned
either parallel or antiparallel.
Using the Fermi's golden rule, we calculate the spin-dependent
tunnel currents across the junctions
\cite{takahashiPRL,takahashiJAP}.
In the following we consider the case that the bias
voltage $\Vb$ is much smaller than the superconducting gap parameter
$\D$.  In this case, the shift of chemical potential $\dmu$ for
up-spin ($-\dmu$ for down-spin) electrons due to spin accumulation
is much smaller than $\D$, so that the tunnel current
$I_{i}^{\s}$ across the $i$th junction ($i=1,2$) becomes
  \begin{eqnarray}                                      
    I_{1}^{\us}(V) = G_{1}^{\us}\Xy \[{\Vb/2} - \dmu/e \],
      \label{eq:I1-u} \\                                
    I_{1}^{\ds}(V) = G_{1}^{\ds}\Xy \[{\Vb/2} + \dmu/e \] ,
      \label{eq:I1-d} \\                                
    I_{2}^{\us}(V) = G_{2}^{\us}\Xy \[{\Vb/2} + \dmu/e \],
      \label{eq:I2-u} \\                                
    I_{2}^{\ds}(V) = G_{2}^{\ds}\Xy \[{\Vb/2} - \dmu/e \].
      \label{eq:I2-d}                                   
  \end{eqnarray}                                        
Here, $G_{i}^{\s}\Xy$ ($i=1,2$) is the tunnel conductance for electrons
with spin $\s$ in the superconducting state, $G_{i}^{\s}$ is that
in the normal state, and
  \begin{equation}              
    \Xy = 2 \int_\D^\infty      
    {\Ek \/ \sqrt{\Ek^2-\D^2}}  
     \(-{\d f_0\/\d\Ek}\) d\Ek, 
     \label{eq:Xy}              
  \end{equation}                
where $f_0(\Ek)$ is the Fermi distribution function and
$\Ek=\sqrt{\xk^2+\D^2}$ the dispersion of quasiparticles, $\xk$ being
one-electron energy relative to the chemical potential.

The spin density $\S$ accumulated in SC is determined by balancing
the spin injection rate $\(d\S/dt\)_{\rm inj}$ with the spin relaxation
rate $\S/\tS$:
  \begin{eqnarray}                              
   \(I_{1\us}-I_{1\ds}\) - \(I_{2\us}-I_{2\ds}\)
         = 2e{ \S / \tS},
      \label{eq:balance}                        
  \end{eqnarray}                                
where $\tS$ is the spin relaxation time and
  \begin{eqnarray}                      
    \S = {1\/2} \sum_{\k}               
     \[f_\us(\Ek) - f_\ds(\Ek)\] \approx N(0)\Xy \dmu,  
     \label{eq:S}                       
  \end{eqnarray}                        
where $f_{\s}(\Ek) \sim f_0(\Ek)-(\d f_0/\d\Ek) \s\dmu$ is the
distribution function of quasiparticles with spin $\s$
and $N(0)$ is the normal-state density of states in SC.

It follows from  Eqs.~(\ref{eq:I1-u})-(\ref{eq:S}) that the tunnel
currents for the parallel (P) and antiparallel (AP) alignments are
given by
  \begin{eqnarray}                                      
    \IF &=&  \Xy {\Vb / \RT},
      \label{eq:IF}             \\
    \IA &=& \[{1-P^2+\GS \/ 1 + \GS }\]  \Xy {\Vb / \RT},
      \label{eq:IA}                                     
  \end{eqnarray}                                        
where $\RT=1/\GT$ ($\GT=G_i^\us+G_i^\ds$) is the tunnel resistance
and $\GS$ is the relaxation parameter
  \begin{eqnarray}              
    \GS = e^2N(0) \RT Ad/\tS,   
      \label{eq:GS}             
  \end{eqnarray}                
with $A$ being the junction area.  
Therefore, we have the TMR ratio at low bias ($\Vb\ll \D$)
  \begin{eqnarray}                      
    TMR = {\IF-\IA \/ \IA} = { P^2 \/ 1-P^2+\GS},       
     \label{eq:TMR}                     
  \end{eqnarray}                        
where $P = {(G_i^\us-G_i^\ds)/(G_i^\us+G_i^\ds)}$ is the tunneling
spin polarization.
For a weak spin relaxation ($\GS \ll 1$), $TMR = P^2 / (1-P^2)$, while
for a strong spin-relaxation ($\GS \gg 1)$, $TMR = P^2 / \GS \ll 1$.


In SC, the spin relaxation is caused by the spin-orbit scattering from
impurities or grain boundaries.  The spin-orbit interaction
$\Hso$ via impurity potential $\Vimp(\r)$ is given by
  \begin{eqnarray}                              
     \Hso=-i (\h/2mc)^2 \vec{\s}\cdot\[\nabla\Vimp(\r) \times \nabla\],
  \end{eqnarray}                                
where $\vec{\s}$ is the Pauli spin matrix.
The scattering matrix elements over quasiparticle states $|\k\s\>$
 has the form:
  \begin{eqnarray}                              
   \<\k'\s'|\Hso|\k\s\> =                       
   i \lso \tilde{V}_{\k'\k}
   \[ {\vec\s}_{\s'\s}\cdot (\hat{\k}\times\hat{\k'}) \],
  \end{eqnarray}                                
where $\lso$ is the spin-orbit coupling parameter,
$\tilde{V}_{\k'\k} = \(u_{\k'}u_{\k} -v_{\k'}v_{\k}\)\Vimp$,
$|u_\k|^2 = 1-|v_\k|^2 = {1\/2}\( 1+{\xk/E_\k}\)$, and
$\hat{\k}=\k/|\k|$.
Using the golden rule formula, we obtain the spin-relaxation rate due to
the spin-flip scattering by $\Hso$:
  \begin{eqnarray}                              
   \( \d\S \/ \dt \)_{\rm sf}                   
    = {2\pi\/\h} \nimp \sum_{\k'\k}     
    |\<\k'\ds|\Hso|\k\us\>|^2 \delta\(\Ek-\Ekp\) \cr \times
        \[f_\ds(\Ekp) - f_\us(\Ek)\]    \nonumber \\
    = - {8\lso^2 N(0)\/9\timp } 
     \int_\D^\infty \[f_\us(\Ek)-f_\ds(\Ek)\] d\Ek,
     \label{eq:dSdt}                            
  \end{eqnarray}                                
where $1/\timp=(2\pi/\h)\nimp\Vimp^2N(0)$ is the scattering
rate by impurities and $\nimp$ is the impurity concentration.

\begin{figure}                                  
  \epsfxsize=0.82\columnwidth                   
  \centerline{\hbox{\epsffile{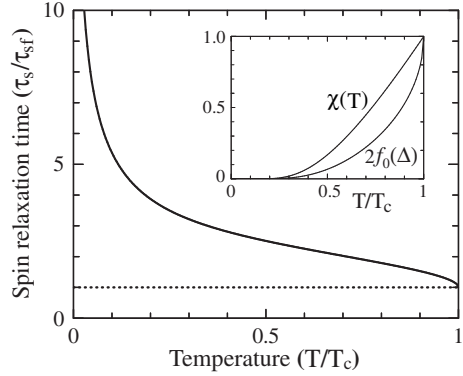}}}       
  \vskip -0.6cm
  \caption{                                     
Temperature dependence of the spin-relaxation   
time $\GSN$. Inset shows $\Xy$ and $2f_0(\D)$
versus $T$, which are used to calculate $\tS$.  
   }                                            
\end{figure}                                    

From Eqs.~(\ref{eq:S}) and (\ref{eq:dSdt}),
we determine the relaxation time $\tS$ from
   $\(\d\S / \dt\)_{\rm sf} = - {\S / \tS}$,
and obtain
  \begin{eqnarray}                              
   \tS                          
   = \tsf {\int_\D^\infty {E\/\sqrt{E^2-\D^2}}  
       \[ f_\us(E) - f_\ds(E) \] dE             
        \/ \int_\D^\infty                       
        \[ f_\us(E) - f_\ds(E) \]dE } ,         
     \label{eq:tS}                              
  \end{eqnarray}                                
where $\tsf=9\timp/8\lso^2$ is the spin-flip scattering time in the normal
state.  Note that the expression of Eq.~(\ref{eq:tS}) is valid for 
$\vF\tsf \gg \xi_0=\h\vF/\pi\D_0$ \cite{maki}.
For $\dmu \ll \D$, Eq.~(\ref{eq:tS}) reduces to
  \begin{eqnarray}                      
    \tS = \[{\Xy/2f_0(\D)}\] \tsf,      
     \label{eq:tS-Yafet}                
  \end{eqnarray}                        
which is the same as the result of Yafet \cite{yafet},
but differs from the result of Zhao and Hershfield \cite{zhao}.
Equation~(\ref{eq:tS}) is a generalization of Yafet to the case
of arbitrary value of $\dmu$.

\begin{figure}                                          
  \epsfxsize=0.76\columnwidth                           
  \centerline{\hbox{\epsffile{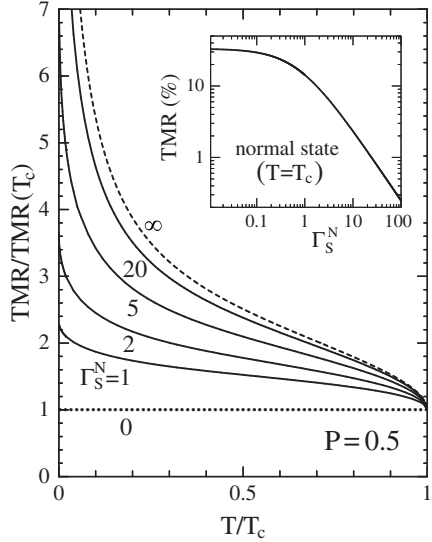}}}               
  \vskip -0.6cm
  \caption{                                             
Tunnel magnetoresistance as a function of temperature   
for different value of the relaxation parameter.  Inset 
shows the TMR versus $\GSN$ in the normal state.        
   }                                                    
\end{figure}                                            


The temperature dependence of the spin-relaxation parameter $\GS$
is scaled to the normalized spin-relaxation time $\tS/\tsf$
by the relation
$\GS = \({\tsf / \tS}\) \GSN$,
where $\GSN=e^2N(0) \RT Ad/\tsf$ is the relaxation parameter in the normal
state.  Figure~1 shows the temperature dependence of $\tS/\tsf$.  Above
$T=T_c$,  
the spin relaxation time
$\tS$ coincides with the spin-flip scattering time $\tsf$.
As temperature $T$ is lowered below $T_c$,
$\tS$ becomes longer with decreasing $T$ and behaves as
$\tS \simeq (\pi\D/2k_BT)^{1/2}\tsf$ at low $T$.

Figure~2 shows the temperature dependence of the normalized TMR for
different values of $\GSN$.  The inset shows the TMR versus $\GSN$ in
the normal state.  In the case of $\GSN >1$, which corresponds to the
case that the spin-relaxation rate is larger than the spin-injection
rate in the normal state, the TMR above $T_c$ is suppressed compared
with the optimal value 33\% for $\GSN=0$ and $P=0.5$ as shown in the
inset of Fig.~2.  However, in the superconducting state below $T_c$,
the TMR increases rapidly with decreasing $T$ due to the increase of
$\tS$, and recovers the optimal TMR in the limit of $T \rightarrow 0$.
If one uses the values of $\RT A=100$~$\W\mu$m$^2$, $\tsf=10^{-10}$sec,
$d=10$ nm, and $N(0)=10^{22}$/(eVcm$^3$), then one obtains $\GSN=10$.
Notice that in the case of strong spin-relaxation ($\GSN \gg 1$),
the TMR becomes proportional to $\tS$, so that the temperature
dependence of $TMR/TMR(T_c)$ coincides with that of $\tS$/$\tsf$
as shown by the dashed curve in  Fig.~2.
The result indicates that the TMR of FM/SC/FM junctions provides
a method to extract important information about spin-relaxation in
superconductors.

\bigskip\noindent

The authors are grateful to A. Fert and M. Johnson for fruitful discussions.
A part of this work was done during stay (S.T.) in CNRS/Thomson-CSF, France.
This work is supported by a Grant-in-Aid for Scientific Research
from Ministry of Education. 
S.M. acknowledges support of the Humboldt Foundation.



\begin{thebibliography}{9}
\small

\bibitem{meservey}
{\it Spin dependent transport in magnetic nanostructures}, edited by
S. Maekawa and T. Shinjo (Gordon and Breach Sci. Pub., London) (in press).

\bibitem{johnson}
M. Johnson and R.~H. Silsbee,
Phys. Rev. Lett. {\bf 55} (1985) 1790.

\bibitem{varet}
T. Varet and A. Fert,
Phys. Rev. B {\bf 48} (1993) 7099.

\bibitem{vanwees}
M. Jemeda {\it et al.},
Nature {\bf 410} (2001) 345.

\bibitem{johnsonS}
M. Johnson, Appl. Phys. Lett. {\bf 65} (1994) 1460.

\bibitem{vasko}
V.~A. Vas'ko {\it et al.}
Phys. Rev. Lett. {\bf 78} (1997) 1134.

\bibitem{dong}
Z.~W. Dong {\it et al.}  
Appl. Phys. Lett. {\bf 71} (1997) 1718.

\bibitem{daibo}
T. Daibo {\it et al.} (unpublished).

\bibitem{vier}
D.C. Vier and S. Schultz,
Phys. Lett. {\bf 98} (1983) 283.

\bibitem{nemes}
N. M. Nemes {\it et al.},
Phys. Rev. B {\bf 61} (2000) 7118.

\bibitem{takahashiPRL}
S. Takahashi, H. Imamura, and S. Maekawa, 
Phys. Rev. Lett. {\bf 82} (1999) 3911.

\bibitem{takahashiJAP}
S. Takahashi {\it et al.}, 
J. Appl. Phys. {\bf 87} (2000) 5227;
{\it ibid.} {\bf 89} (2001) 7505.

\bibitem{takahashiPC}
S. Takahashi, H. Imamura, and S. Maekawa,
Physica C {\bf 341-348} (2000) 1515.

\bibitem{maki}
K. Maki, Phys. Rev. B {\bf 8} (1973) 191;
L. R. Tagirov {\it et al.}, J. Phys. F {\bf 17} (1987) 695.

\bibitem{yafet}
Y. Yafet, Phys. Lett. A {\bf 98} (1983) 287.

\bibitem{zhao}
H.~L. Zhao and S. Hershfield,  
Phys. Rev. B {\bf 52} (1995) 3632.
In their result, $\Xy$ is missing in Eq.~(\ref{eq:tS-Yafet}).
\end{thebibliography}
\end{document}